\documentclass[11pt]{iopart}
\usepackage[utf8]{inputenc}
\usepackage{multicol}
\usepackage{multirow}
\usepackage{booktabs} 
\usepackage{float} 
\usepackage{hyperref} 
\usepackage{paralist} 

\usepackage{capt-of}

\usepackage{paralist}
\usepackage{multicol}
\usepackage{setspace}
\usepackage{graphicx}
\usepackage{multirow}
\usepackage[table]{xcolor}
\usepackage{graphicx}
\usepackage{makeidx}
\usepackage{subfigure}
\usepackage{geometry}
\usepackage{xcolor}
\usepackage{algorithm2e}
\usepackage{pifont}
\newcommand{\boldsymbol}[1]{\mathbf{#1}}
\definecolor{light-gray}{gray}{0.9}

\newcommand{\fmin}{\ensuremath{f_\mathrm{min}}}
\newcommand{\fmax}{\ensuremath{f_\mathrm{max}}}
\newcommand{\pvec}{\ensuremath{\vec{\theta}}}
\newcommand{\data}{\ensuremath{\vec{d}}}
\newcommand{\df}{\ensuremath{\delta f}}
\newcommand{\dt}{\ensuremath{\delta t}}
\newcommand{\Nfix}{\ensuremath{N_\mathrm{fix}}}

\newcommand{\Msun}{\ensuremath{M_\odot}}

\begin{document}

\title{Accelerating gravitational wave parameter estimation with multi-band template interpolation}

\author{Serena~Vinciguerra, John~Veitch, Ilya~Mandel}
\address{Institute of Gravitational Wave Astronomy \& School of Physics and Astronomy, University of Birmingham, Birmingham, B15 2TT, United Kingdom}
\ead{serena@star.sr.bham.ac.uk}

\date{\today}


\begin{abstract}
Parameter estimation on gravitational wave signals from compact binary
coalescence (CBC) requires the evaluation of
computationally intensive waveform models, typically the bottleneck in the
analysis. This cost will increase further as low frequency sensitivity in later second and third generation detectors motivates the use of longer waveforms.

We describe a method for accelerating parameter estimation by exploiting the
chirping behaviour of the signals to sample the waveform sparsely for portions
where the full frequency resolution is not required. We demonstrate that the method can reproduce the original results with a waveform mismatch of $\leq 5\times 10^{-7}$, but with a waveform generation cost up to {$\sim\,50$} times lower for computationally costly frequency-domain waveforms starting from below 8\,Hz.
\end{abstract}

\maketitle 

\section{Introduction}
The discovery of gravitational waves from coalescing binary black hole systems
made by Advanced LIGO in its first observing run opened the door to  gravitational wave astronomy~\cite{GW150914-DETECTION}.
As the second generation of ground based
detectors continues to evolve towards their design sensitivities the rate of
detections is expected to increase, leading eventually to the detection of
lower mass binary systems such as binary neutron star (BNS) and neutron star -
black hole (NSBH) binaries~\cite{abadie2010predictions}.

The characterisation of these sources involves the use of Bayesian parameter
estimation and model selection algorithms based on stochastic sampling of the
posterior probability distribution for the model parameters conditioned on the
observed data. This process involves repeated comparisons of the data with
template waveforms through evaluation of the likelihood function. Previous
implementations (e.g. \texttt{LALInference}~\cite{Veitch:2015}) have required millions of likelihood evaluations, which implies that a similar
number of template waveforms must be generated. In the case of sophisticated
waveform models this template generation dominates the computational cost of
the analysis, with the cost scaling linearly with the length of the waveform
$\tau$, which in turn scales with the low frequency starting point of the waveform as
$\fmin^{-8/3}$. As the low-frequency sensitivity of the second-generation
instruments improves, \fmin\ is expected to reduce from $\sim 30$\,Hz to $\sim
10$\,Hz or lower. The issue becomes even greater in the case of
subterranean third-generation instruments such as the Einstein Telescope which
are expected to reduce this further to $5$\,Hz or lower~\cite{ETDesign2011,Punturo:2010}.
This improvement in low-frequency sensitivity should translate to much more accurate estimation of key parameters. However, taking full advantage of this improvement in a timely and computationally efficient manner is a challenge. We present a method that leverages the frequency evolution of the waveform to effectively reduce the number of waveform samples that must be computed. This has the potential to asymptotically reduce the computational cost of template generation by a factor that is proportional to $\fmin^{-1}$. Here we give details of a practical implementation which does not compromise the accuracy of parameter estimation and study the computational cost scaling in a realistic analysis.

Several methods have been developed previously to overcome the need to
evaluate the waveform and likelihood at each point in the Fourier domain.
Reduced order quadrature (ROQ) methods, first introduced for CBC waveforms in
\cite{Field:2011} and developed for the purpose of parameter estimation (PE) in
\cite{Canizares:2013,Smith:2013,Canizares:2014,Smith:2016qas}, seek to represent the waveform in an alternative basis
from the standard Fourier components. A waveform for a particular point in
parameter space is represented as the linear combination of a number of these
basis templates. By projecting the data into the same basis the likelihood
function can be computed using a sum over bases rather than a sum over Fourier
components, where the number of bases is far smaller than the number of
Fourier components. This method significantly accelerates the
likelihood computation.  However, it has the drawback of
requiring the basis to be constructed in advance for each waveform  family, a process
which is costly in terms of both computation and memory requirements to store
the input waveforms, with a cost that grows rapidly as the dimensionality
of the model is increased to include misaligned spins.
The large intrinsic volume of the mass parameter space requires that it be
subdivided into patches of manageable size, with each patch having a different
set of bases.
The ROQ likelihood calculation is also dependent on the particular noise curve used through the ROQ integration weights, which must be computed for the particular characteristics of the data at the time of the event of interest.
Furthermore, severing the link between frequency and the representation of the waveform makes it difficult to model the effect of frequency-dependent detector calibration
errors, which were included in the analysis of binary black hole systems in
O1~\cite{GW150914-PARAMESTIM,CBC-O1-BBH}.

A different approach has been developed in the context of low-latency searches
for gravitational waves. In this context the incoming data-stream is filtered
against a pre-determined bank of templates which is chosen to cover the
mass parameter space with a certain maximum guaranteed loss of signal-to-noise
ratio (SNR). Although here
the filtering can proceed in parallel it is still desirable to reduce the cost
of the search by reducing the volume of data that has to be processed. The
MBTA \cite{abadie2012first,adams2015low} and gstlal~\cite{Cannon:2012} pipelines
divide the templates into bands, which are chosen
to exploit the chirping nature of the inspiral signal. Each band  has a certain
maximum signal frequency $f<\fmax$, so both the template and the data can be
down-sampled to a lower sampling rate, reducing the cost of the filtering
process for each band. The original high-bandwidth SNR time-series can be
reconstructed from the output of the banded filters by subsequent
up-sampling, which can be done selectively on data stretches which have
significant SNR in the banded filters.
A similar approach has been advocated for LISA data analysis, employing two bands
for each template~\cite{porter2014new}, as is also the case for MBTA.

In this paper we pursue an approach inspired by the latter method of
subdividing the waveform into band-limited pieces, with the aim of using it
for PE rather than searching. This places some additional
constraints on the accuracy of waveform reconstruction required to
reproduce the results from a full-bandwidth analysis without adding
systematic or statistical errors. Our method is currently limited to the
computation of the template (likelihood evaluation is still performed in the
full Fourier basis), but nevertheless can produce large reductions in
computational cost for long duration signals when the more sophisticated (and
costly) waveform models are employed. Unlike the ROQ, this allows us to
maintain the link with frequency and easily include calibration error modelling in
the analysis. Also, because the method requires no pre-computation of a new
basis it can be applied without modification to any frequency-domain waveform
model, including modifications to the signal such as tidal
effects~\cite{vines2011post,hinderer2016effects} and
parameterised deviations from general relativity~\cite{GW150914-TESTOFGR,agathos2014tiger}.
This flexibility is the main advantage of the method, which makes it
especially suitable for analyses where an ROQ model is not available, or where
its production would be too costly. An implementation is
provided in the open source \texttt{LALInference} PE
software~\cite{Veitch:2015}. We describe the method in detail in section
\ref{s:method} and demonstrate its efficacy when applied to the  analysis of
simulated signals in section \ref{s:results}.  We discuss possible future developments in section \ref{s:conclusion}.


\section{Multi-banding approach: the method}\label{s:method}
\label{method}

\subsection{Motivation}
In gravitational wave PE, the aim is to explore the
posterior probability distribution of the source model,
\begin{equation}
        p(\pvec|\data,H) = \frac{p(\pvec|H)p(\data|\pvec,H)}{p(\data|H)}
\end{equation}
where $\pvec$ are the physical parameters of the source such
as the masses, spins, position and orientation~\cite{Veitch:2015}. The likelihood function for a single detector under the
assumption of Gaussian noise depends on the data $\data$ and the parameter $\pvec$,
as well as the particular waveform model used $H$, as
\begin{equation}
        p(\data|\pvec,H) \propto \exp\left[-2\sum_i^{N}
        \frac{|h_i(\pvec)-\data_i|^2}{\tau S_n(f_i)}\right]
\end{equation}
where $S_n(f_i)$ is the power spectral density of the detector, 
$\tau=\df^{-1}$
is the
duration of the data segment to be analysed, and 
$N= \tau/(2\dt)$
is the number of
Fourier components in the frequency-domain complex representation of the modeled signal
$h_i(\pvec)$ as it would be observed in the detector. Since the details of the detector
responses are not important for what follows we refer the reader to
\cite{Veitch:2015} for a full description of how the extrinsic parameters are
used to construct the observed signal in each detector.
In order to accurately capture the waveform we must choose 
$\tau$
and $\dt$ such that the
entire signal duration, from the time it enters the sensitive band of the
instrument at frequency $\fmin$, is contained in 
$\tau$,
and the sampling resolution $\dt<(2\fmax)^{-1}$
is sufficient to capture the highest frequency components of the signal at \fmax.

\begin{figure}
        \begin{centering}
        \includegraphics[width=\textwidth]{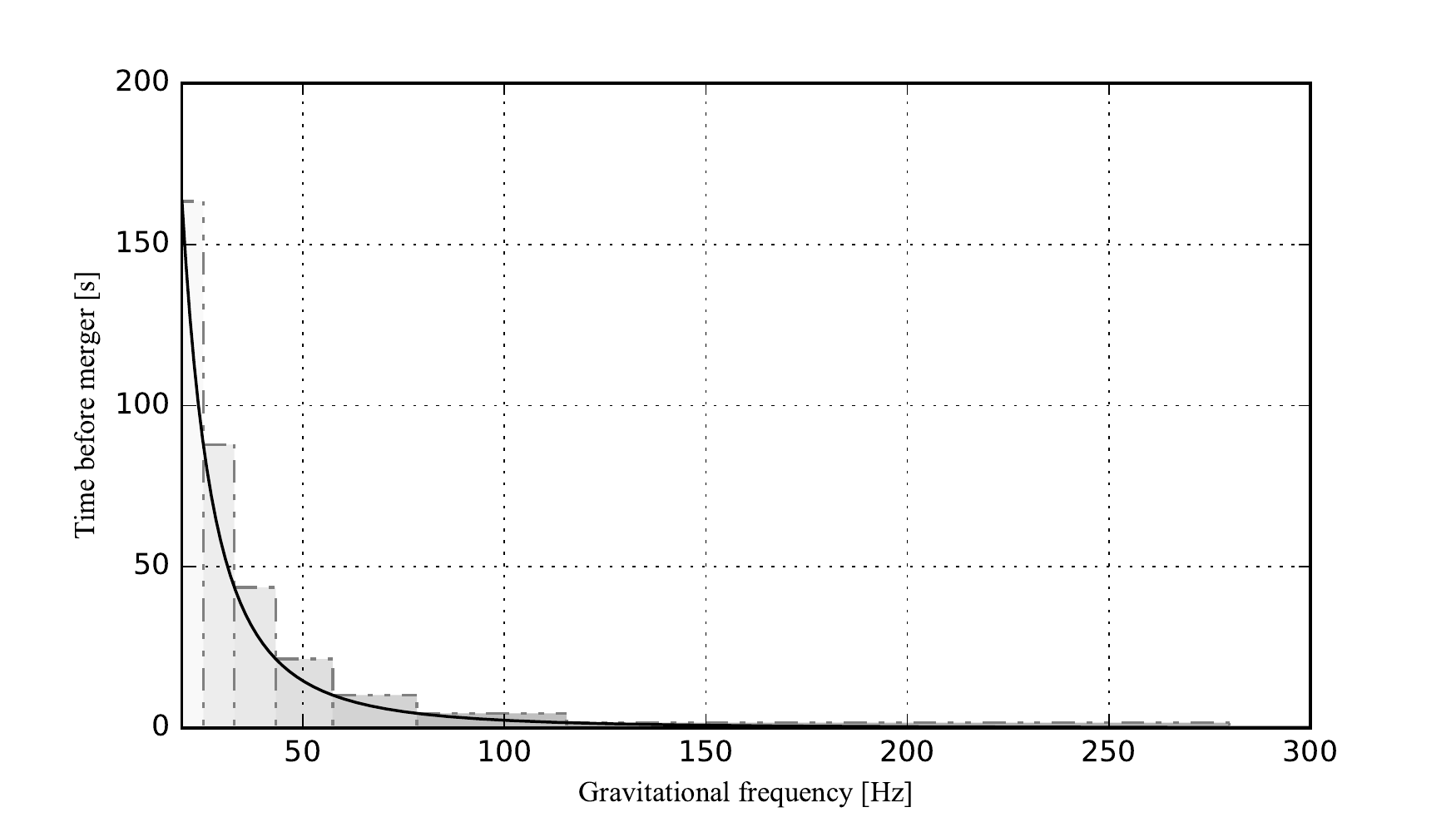}
\caption{Time from a given frequency to coalescence for a fiducial binary neutron star signal. Coloured boxes indicate the subdivision of the waveform into bands with adaptive frequency resolution, as determined by the time before coalescence; see section \ref{Alg0_Fdef}.}\label{fig:TofF}
\end{centering}
\end{figure}

To leading order, the duration of an inspiral signal from a certain frequency
$f$ to the formal time of coalescence is~\cite{Cutler:1994} (in geometrical units $G=c=1$)
\begin{equation}\label{eq:T}
        t(f) \approx 5\left[8\pi f\right]^{-8/3}\mathcal{M}^{-5/3},
\end{equation}
where $\mathcal{M} = M_1^{3/5} M_2^{3/5} (M_1+M_2)^{-1/5}$ is the chirp mass of a binary with component masses $M_{1,2}$ and mass ratio $q=M_2/M_1\leq 1$.  During the inspiral, the gravitational-wave frequency monotonically increases until the merger and
ring-down phases. An example is shown in figure \ref{fig:TofF}, where we put
frequency on the abscissa to emphasize that we are working in the frequency
domain.

In the standard calculation, there is a fixed frequency resolution of 
$\df=\tau^{-1}$
between
frequency bins, and the total number of frequency-domain samples required to describe the signal is
\begin{eqnarray}
        \Nfix & = \int_{\fmin}^{\fmax} \df^{-1} df \nonumber \\
           & = (\fmax - \fmin)\tau \nonumber \\
          & \approx 5(8\pi)^{-8/3} \mathcal{M}^{-5/3}(\fmax-\fmin)\fmin^{-8/3}\, .
\label{eq:fixN}
\end{eqnarray}
We can see from the figure that this frequency resolution is necessary to contain
the full length waveform starting at time $\tau$ before merger, but as the
frequency increases the time before merger $t(f)$ decreases and the waveform
is over-sampled in frequency. Our aim is to take advantage of this to increase
$\df$ as a function of frequency without losing any information about the
waveform phasing, thereby reducing the total number of points at which the waveform
must be evaluated.

We now consider the asymptotic limit of multi-banding.  In the idealized limit, the frequency step $\df=t(f)^{-1}$ can vary continuously throughout the
signal. We then have
\begin{eqnarray}
        N_\mathrm{min} &= \int_{\fmin}^{\fmax} t(f) df \nonumber \\
                       &= -3(8\pi)^{-8/3}\mathcal{M}^{-5/3}(\fmax^{-5/3}-\fmin^{-5/3}) \, .
\label{eq:idealN}
\end{eqnarray}
The relative number of points required for the standard case compared to the ideal case is then
\begin{eqnarray}
        \frac{\Nfix}{N_\mathrm{min}} =
        \frac{5}{3}\frac{(\fmax-\fmin)\fmin^{-8/3}}{\fmin^{-5/3}-\fmax^{-5/3}} \, ,
\label{eq:idealG}
\end{eqnarray}
which for $\fmax \gg \fmin$ indicates an asymptotic reduction in number of
points $5\fmax/3\fmin$. For a binary neutron star waveform which enters the
detector at 20\,Hz and terminates at 1500\,Hz, the potential reduction in
number of points is therefore a factor of $\sim 125$.

\subsection{Choice of bands}
\label{Alg0_Fdef}
Rather than taking the continuously varying \df\ case, in our practical
implementation we work with a pre-determined set of frequencies which
divides the total frequency span into several bands with constant \df\ within
each band.  We position the bands in frequency space so that \df\ changes by a factor of 2 between neighboring bands, while ensuring that the Nyquist sampling criterion is always met.

Figure~\ref{fig:TofF} shows a schematic of the basic idea. We must choose our
bands such that they are able to accurately represent the longest waveform in
our allowed mass prior. This can be determined automatically at run-time of the
PE code; e.g., a $1+1$\Msun\ binary neutron
star signals lasts $281$\,s from $20$\,Hz to coalescence.
Starting at the lowest frequency \fmin, the frequency resolution necessary to
contain the waveform is $\df_0\le t(\fmin)^{-1}$. Each subsequent band has a
sampling rate $\df_b=2\df_{b-1}$ and so the time at the start of the new band is
a factor of two closer to coalescence, $t(f_b)=t(f_{b-1})/2$.
The frequencies at which to place the band boundaries are then determined by
inverting Eq.~\ref{eq:T} and solving for the series of $\df_b$.
To summarise, we can specify the frequencies at which the waveform is evaluated
via the following algorithm\\
\begin{algorithm}[H]
        $b=0$, $i=0$\\
        $\df_b = t(\fmin)^{-1}$, $f_i = \fmin$\\
        \While{$f_i < \fmax$}
        {
                \While{$t(f_i)>(2\df_b)^{-1}$}
                {
                    $f_{i+1}=f_i + \df_b$\\
                    $i=i+1$
                }
                $\df_{b+1}=2\df_b$\\
                $b=b+1$                
        }

\end{algorithm}

\subsection{Up-sampled waveform}\label{ss:Interpolation}
Having defined the reduced set of frequencies at which the waveform is to be
calculated, we now outline the procedure for reconstructing the full waveform.
Note that unlike in reduced order quadrature methods, we still compute the
likelihood using the fully sampled dataset.  A na\"ive decimation
or averaging of the frequency-domain detector data leads to a loss of information
relative to the fully-sampled results. We therefore use an interpolation scheme
to reconstruct the waveform at the full sampling rate in order to match filter the original data.

Direct linear interpolation of the reduced waveform $\tilde{h}(f_j)$ does not
accurately reproduce the original waveform as the oscillatory behaviour is not
captured by the interpolating straight line segments. We therefore work with
the waveform represented in amplitude and phase as $\tilde{h}(f_j)=A_j\exp(i\phi_j)$,
where $j$ labels the reduced set of frequencies.
Within each coarse bin, we linearly interpolate the amplitude $A$ and phase $\phi$ to obtain estimates of the amplitude $\hat{A}_k=\hat{A}(\hat{f}_k)$ and phase $\hat{\phi}_k=\hat{\phi}(\hat{f}_k)$ at the dense set of frequencies $\hat{f}$ labeled with $k$:
\begin{eqnarray}
\hat{A}_k&=A_j+\frac{\hat{f}_k-f_j}{f_{j+1}-f_j}(A_{j+1}-A_j)\, ,\\
\hat{\phi}_k&=\phi_j+ \frac{\hat{f}_k-f_j}{f_{j+1}-f_j}(\phi_{j+1}-\phi_j)\, ,
\end{eqnarray}
where $f_j$ is the nearest coarse frequency point below $\hat{f}_k$ and $\hat{f}_{k+1} - \hat{f}_{k} = \delta f_0$.  
The up-sampled waveform after multi-banding and interpolation (hereafter MB-Interpolation)
is then $\tilde{h}(\hat{f}_k)=\hat{A}_k\exp(i\hat{\phi}_k)$.

One practical problem with applying this formula is that the exact estimation of $\exp{i\hat{\phi}_k}$ is computationally expensive. To avoid this we use the recursive property
$e^{i \hat{\phi}_{k+1}} =e^{i \hat{\phi}_k}e^{i \delta f_0 (\phi_{j+1}-\phi_j)/(f_{j+1}-f_j)}$.  The last term needs to be computed only once for each coarse bin~\cite{press2007numerical}.
The recursion relation can be expressed in terms of the real and imaginary parts of the complex frequency-domain signal as
\begin{eqnarray}
\Re(\hat{h}_{k+1})&=&\left[1+\frac{(A_{j+1}-A_j)\delta f_0}{\hat{A}_k (f_{j+1}-f_j)}\right]\left[\Re(\hat{h}_k)\left(1-2\sin^2\frac{\delta\phi_j}{2}\right)-\Im(\hat{h}_k)\sin{\delta\phi_j}\right] ,\nonumber \\
\Im(\hat{h}_{k+1})&=&\left[1+\frac{(A_{j+1}-A_j)\delta f_0}{\hat{A}_k (f_{j+1}-f_j)}\right]\left[\Im(\hat{h}_k)\left(1-2\sin^2\frac{\delta\phi_j}{2}\right)+\Re(\hat{h}_k)\sin{\delta\phi_j}\right] \nonumber ,
\end{eqnarray}
where $\delta \phi_j \equiv \delta f_0 (\phi_{j+1}-\phi_j)/(f_{j+1}-f_j)$;
therefore we only need to compute $\sin (\delta \phi_j)$ and $\sin^2(\delta \phi_j /2)$.

\subsection{Accuracy}

The waveform accuracy required for parameter estimation is determined by the condition that systematic bias in parameter estimates from imperfect waveforms should be much smaller than the statistical measurement uncertainty of inference on data with finite signal-to-noise ratios (e.g., \cite{Ohme:2012}).  Therefore, the shift in the log likelihood due to the use of MB-Interpolation waveforms in lieu of the original waveforms, $\delta \log L_\textrm{MB-Interpolation}$, should be smaller than the spread in the log likelihood over the posterior $\sigma_{\log L}$:
\begin{equation}
\delta \log L_\textrm{MB-Interpolation} \ll \sigma_{\log L} \sim \sqrt{\frac{N_\textrm{param}}{2}}\, ,
\end{equation}
where $N_\textrm{param}$ is the number of parameters in the model.  This condition on the log likelihood can be expressed in terms of the match between the original waveform $h_0$ and the MB-Interpolation waveform $h$ \cite{baird2013degeneracy,hannam2010length,Haster:2015hda}:
\begin{equation}
\frac{\langle h_0 - h| h_0 -h\rangle}{\langle h_0 | h_0 \rangle} \ll \frac{\sqrt{2N_\textrm{param}}}{\rho^2}\, ,
\end{equation}
where $\rho$ is the signal-to-noise ratio.  Considering $\rho\sim 20$, typical for a moderately loud signal \cite{chen2014loudest}, the threshold on the mismatch is $\sim 10^{-3}$.  

\begin{figure}[!hbt]
{\includegraphics[width=16.9cm]{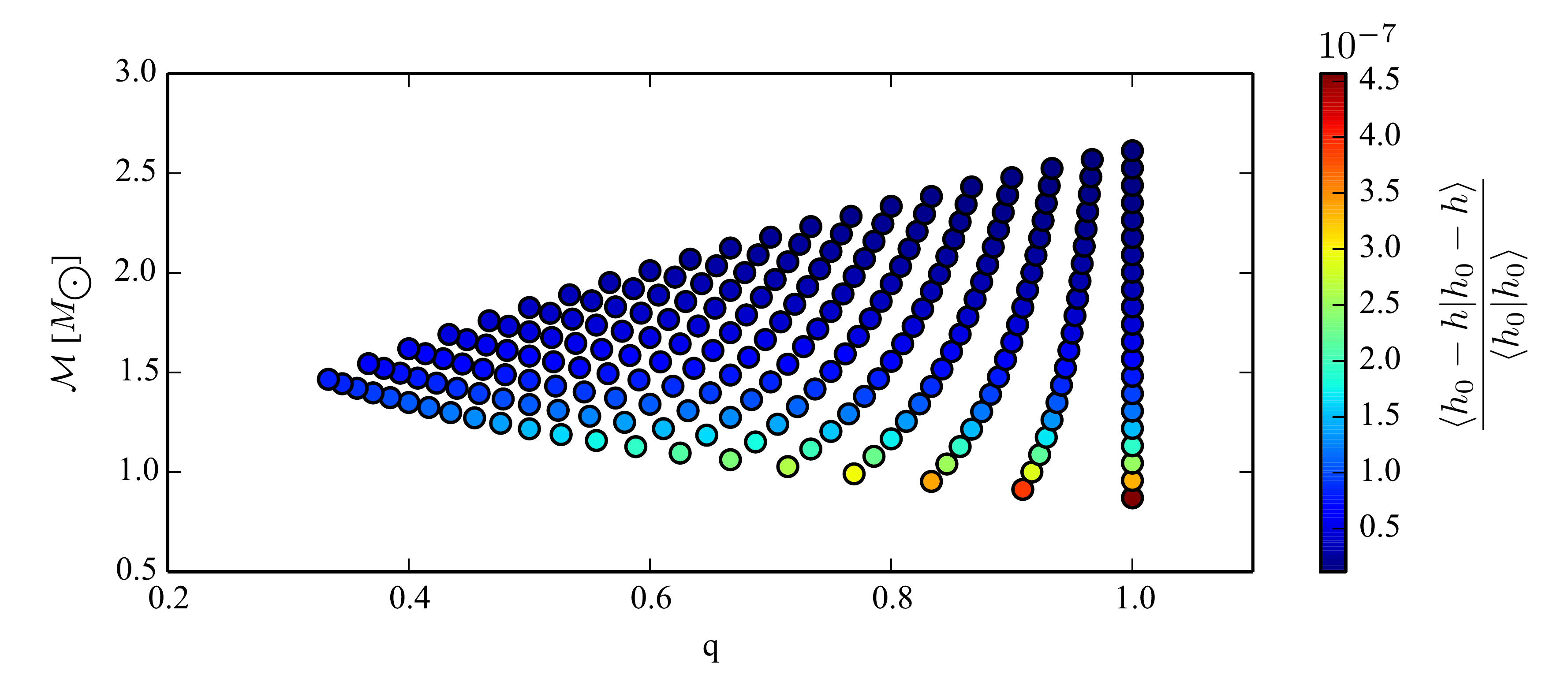}}
\caption{Mismatch of MB-Interpolation waveforms ($h$) against waveforms computed with the standard procedure ($h_\mathrm{0}$) as a function of chirp mass and mass ratio. The mismatch is calculated up to 1024Hz.
}
\label{fig:accuracy}
\end{figure}

Figure \ref{fig:accuracy} show that the mismatch of MB-Interpolation waveforms against the original waveforms is a factor of a thousand smaller than this requirement over the binary neutron star region, decreasing for more massive systems. This is expected, as in the frequency domain the density of cycles at low frequency increases with the time duration of the waveform, so the most demanding case is that of the lowest mass considered in a particular analysis (in our case a $1-1\,\Msun$ binary). Therefore, we conclude that this procedure provides sufficient accuracy for unbiased inference at all masses above $1-1\,\Msun$.

\section{RESULTS}\label{s:results}


We  implemented the MB-Interpolation approach (section \ref{s:method}), including the waveform interpolation procedure (subsection \ref{ss:Interpolation})
within \texttt{LALInference} \cite{Veitch:2015}.
We performed several tests in order to validate MB-Interpolation.  We first checked the effectiveness of the MB-Interpolation by verifying the reduction of the number of frequencies at which the template is evaluated when multibanding. We then measured the speedup in the waveform computation following multibanding and interpolation.  Finally, we tested the 
overall acceleration of the complete PE analysis with MB-Interpolation and confirmed its accuracy.

\subsection{Reduction of template evaluations}
\begin{figure}
    \centering
\includegraphics[]{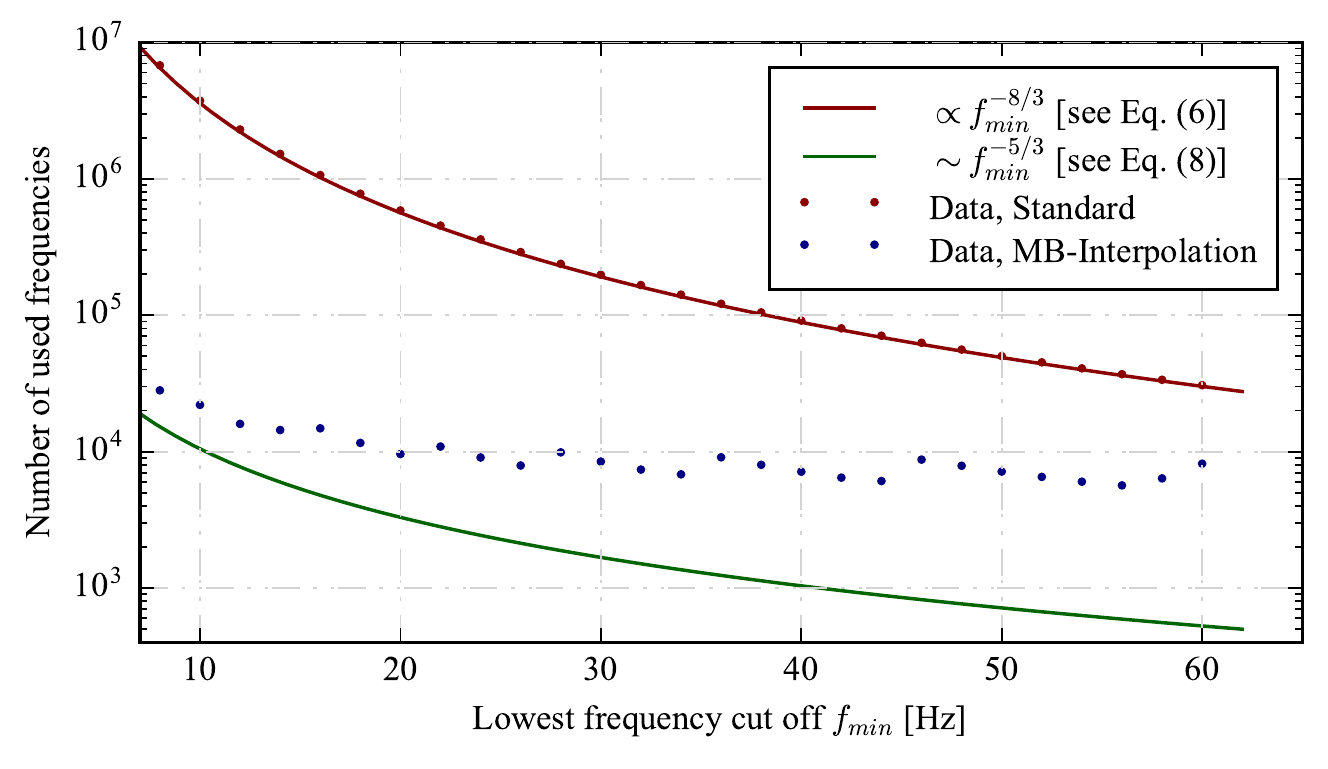}
\caption{
Number of frequencies at which the waveform is evaluated when using the standard ($\Nfix$, red dots) and MB-Interpolation ($N_\mathrm{MB}$, blue dots) algorithms as a function of the lower frequency limit. 
The red curve corresponds to equation \ref{eq:fixN} while the green curve shows the number of frequency samples in the theoretical limit of continuously adapted sampling steps, equation \ref{eq:idealN}. MB-Interpolation is sub-optimal but approaches the asymptotic case in the limit $\fmin\rightarrow 0$, as the templates become very long and $\df_0$ approaches 0; the number of frequency bands increases from 3 at $\fmin = 60$ Hz to 8 at $\fmin = 20$ Hz and 11 at $\fmin = 8$ Hz.
}
    \label{fig3}
\end{figure}

To measure the speedup in waveform generation we first defined the frequency set at which the multibanded template is evaluated   
according to the algorithm in section  \ref{Alg0_Fdef}. The in-band signal duration is set by a BNS with both component masses equal to $1M_{\odot}$ as a reference system, corresponding to the lowest limit of the component mass prior adopted in the analysis.
The number of frequencies at which the waveform is evaluated is shown in figure \ref{fig3} as a function of the starting frequency $\fmin$ for both MB-Interpolation and the standard algorithm.
This figure clearly demonstrates the effectiveness of the approach in reducing 
template evaluations: 
the number of frequencies defining the two sets, $\Nfix$ and $N_\mathrm{MB}$ respectively for the standard and the MB-Interpolation algorithm, differs by an order of magnitude or more for starting frequencies below $40$ Hz.

The evident segmented structure of $N_\mathrm{MB}$ reflects the varying number of frequency bands used in MB-Interpolation.  Within each band, \df\ is constant and the number of frequencies follows the same $\sim \fmin^{-8/3}$ scaling as for the standard algorithm.
As expected, this yields sub-optimal behavior relative to the theoretical limit of a continuously varied sampling frequency, as clearly demonstrated by the ideal case (green line) falling well below the actual $N_\mathrm{MB}$ points in the same figure.

\subsection{Speedup of waveform generation}
\label{subs:WG}
We measured the reduction in the total waveform generation time, including both multibanding and interpolation, for compact binary systems with chirp-mass of $\sim 1.48$\,\Msun.
The waveforms were generated up to a frequency $\fmax$ of 2048\,Hz with a time domain sampling rate of 4096\,Hz.
We used two different waveform models for both generating and analysing injections to test the efficacy of our approach: {\it TaylorF2} (see for example \cite{buonanno2009comparison}) and {\it IMRPhenomPv2} \cite{hannam2014simple}. The former is one of the simplest and most common waveform models available for the coalescence of compact binaries. It analytically describes the inspiral stage of the coalescence using the {\it stationary phase approximation}. Meanwhile, the analytical {\it IMRPhenomPv2} model includes the inspiral, merger and  ringdown phases, calibrated to numerical relativity simulations.
The {\it IMRPhenomPv2} waveform family has been 
used to characterize the BBH systems discovered during O1, the first science run of Advanced LIGO \cite{CBC-O1-BBH}.
{\it IMRPhenomPv2} waveforms are more sophisticated and more computationally expensive than {\it TaylorF2} ones. Since the main effect of the proposed method is reducing the number of template evaluations, it is for computationally expensive cases that we expect to benefit the most from its application.

\begin{figure}[t]
\mbox{
\subfigure{\includegraphics[]{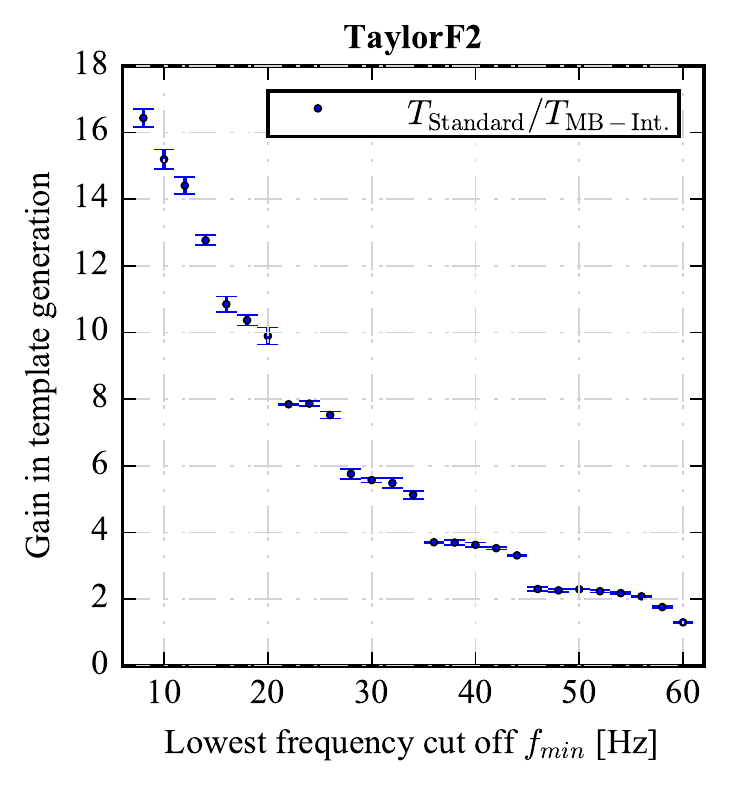}}\,
\subfigure{\includegraphics[]{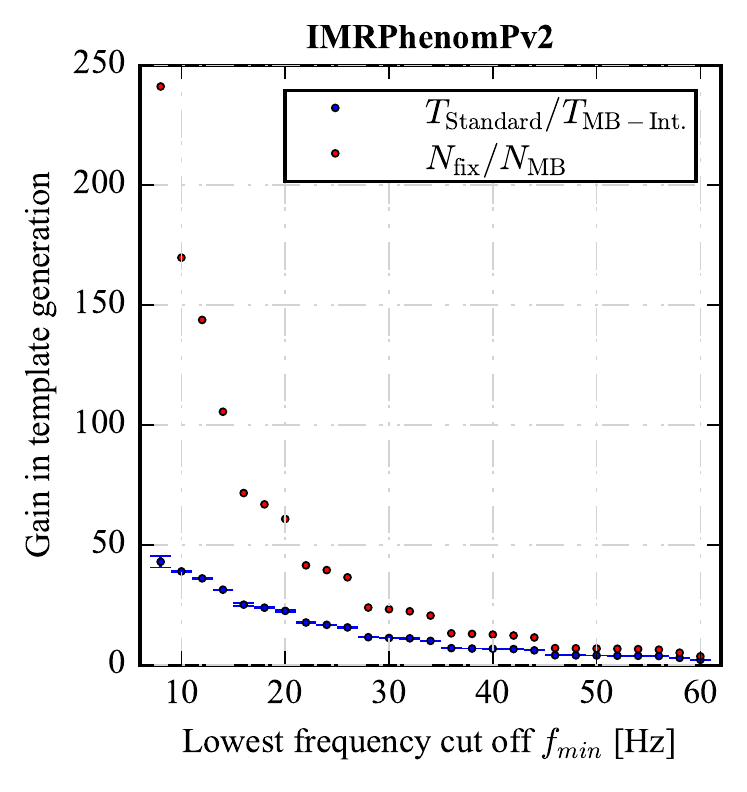}}
}
\vspace{-0.6 cm}
\caption{
Gain factor in computational speed of template generation as a function of $\fmin$.
Blue points: ratio of the average waveform computation cost for the standard procedure versus MB-Interpolation as a function of the starting frequency for {\it TaylorF2} (left panel) and {\it IMRPhenomPv2} (right panel) waveform families. Red points in right panel: ratio between the number of frequencies at which the waveform is evaluated when using the standard procedure versus MB-Interpolation. 
}
\label{fig2_2}
\end{figure}

Figure \ref{fig2_2} shows the speedup in the template generation as a function of the starting frequency, for the {\it TaylorF2} waveform model in the left panel, and for {\it IMRPhenomPv2} in the right one.
The length of the data segments was set by calculating the duration of a BNS signal with $1\,M_\odot$ components starting from the chosen $\fmin$.
The template generation speed was calculated by averaging the time necessary to construct one waveform over 3000 (300) trials for the {\it TaylorF2} ({\it IMRPhenomPv2}) model. We define the gain in speed (blue points in figure \ref{fig2_2}) as the ratio between the average time required by the standard and the MB-Interpolation algorithms to compute one template. 

For comparison, the right panel of figure \ref{fig2_2} includes the reduction in the number of frequencies at which the waveform is evaluated when using MB-Interpolation, $\Nfix/N_\mathrm{MB}$ (red points).
The gains for MB-Interpolation are smaller than the ratio $\Nfix/N_\mathrm{MB}$ because of the additional cost of interpolating between the $N_\mathrm{MB}$ frequency samples.

We find that the MB-Interpolation scheme yields a dramatic gain in computational speed for smaller values of \fmin. At $\fmin=20$\,Hz {\it TaylorF2} templates were accelerated by a factor of 10.  The slower {\it IMRPhenomPv2} family shows significantly greater gains than the faster {\it TaylorF2} family, as illustrated by the difference in ordinate scales between the two panels of  figure \ref{fig2_2}.  Thus, {\it IMRPhenomPv2} template generation was around 25 times faster with MB-Interpolation at $\fmin=20$\,Hz. While both waveform families show a greater gain for smaller values of $\fmin$, the same segmentation is present as in figure \ref{fig3}, although this is less obvious for the {\it TaylorF2} family. This reflects the dependence of the time required to compute one waveform, $T$,
on the number of frequency bands.

Within the standard approach, this time $T_\mathrm{Standard}$ 
can be approximated as the product of the time necessary to calculate the waveform at a given frequency $t^\mathrm{w}$ with the number frequencies $\Nfix$:
\begin{equation}
T_\mathrm{Standard} \sim \Nfix \cdot \, t^\mathrm{w}.
\end{equation}

To estimate the same time in the MB-Interpolation algorithm we need to take into account two different contributions: the template generation applied to a reduced set of frequencies, and the calculations necessary for the waveform interpolation $t^\mathrm{i}$. This leads to the following approximation:
\begin{eqnarray}
T_\mathrm{MB-Int.} \sim \Nfix\cdot t^\mathrm{i} + N_\mathrm{MB}\cdot \left(t^\mathrm{w} + \delta t^\mathrm{i}\right) \, 
\label{eqA}
\end{eqnarray}
Here 
$\delta t^\mathrm{i}$ represents the time required to compute the quantities necessary for the interpolation (such as phase, derivatives, etc.); typically $\delta t^\mathrm{i}\ll t^\mathrm{w}$.

According to Eq.~\ref{eqA}, the time required to compute a complete waveform via MB-Interpolation depends on $\fmin$ only through $\Nfix$ and $N_\mathrm{MB}$. However, the first term in Eq.~\ref{eqA} becomes increasingly dominant as $\fmin$ decreases, since $\Nfix\propto \fmin^{-8/3}$.  For sufficiently small $\fmin$, $\Nfix\cdot t^\mathrm{i} \gg N_\mathrm{MB}\cdot \left(t^\mathrm{w} + \delta t^\mathrm{i}\right)$ and the speedup asymptotes to a fixed factor 
$T_\mathrm{Standard} / T_\mathrm{MB-Int.} \to  t^\mathrm{w}/ t^\mathrm{i}$,
independent of $\fmin$. The frequency at which this happens depends in general on the computational cost of the particular waveform model. The results reported in figure \ref{fig2_2} suggest gains exceeding $\sim 16$ ($\sim 50$) for starting frequencies  below $8$ Hz for the {\it TaylorF2} ({\it IMRPhenomPv2}) waveform models.

\subsection{Inference} 
To verify that the results obtained with the MB-Interpolation algorithm remain accurate, and to measure the speedup of an end-to-end inference run,
we also performed several complete PE analyses.
We injected a gravitational wave signal emitted by a neutron star binary with component masses $M_1 = M_2 = 1.4$\,\Msun\ into stationary Gaussian noise, coloured according to the design sensitivity curves of advanced LIGO and Virgo \cite{aasi2016prospects}.
The signal was always injected at a distance of $D_L\approx 200$\,Mpc so that the SNR at  $\fmin=40$\,Hz source was 15; signals with lower \fmin\ have correspondingly higher SNR.

The PE analyses were performed with \texttt{LALInferenceNest}, using the same maximum frequency $\fmax = 2048$\,Hz and time-domain sampling rate (4096\,Hz) adopted in section \ref{subs:WG}. 
Priors on companion masses were uniform in the range $1\-- 3$\,\Msun\ and the prior on distance was uniform in volume with a maximum distance of 500\,Mpc.
We chose this region of mass space as it is the most challenging in terms of computational cost and has the strictest accuracy requirements for waveform interpolation.

\subsubsection{PE Consistency}
\begin{figure} [ht]
 \centering
\includegraphics[]{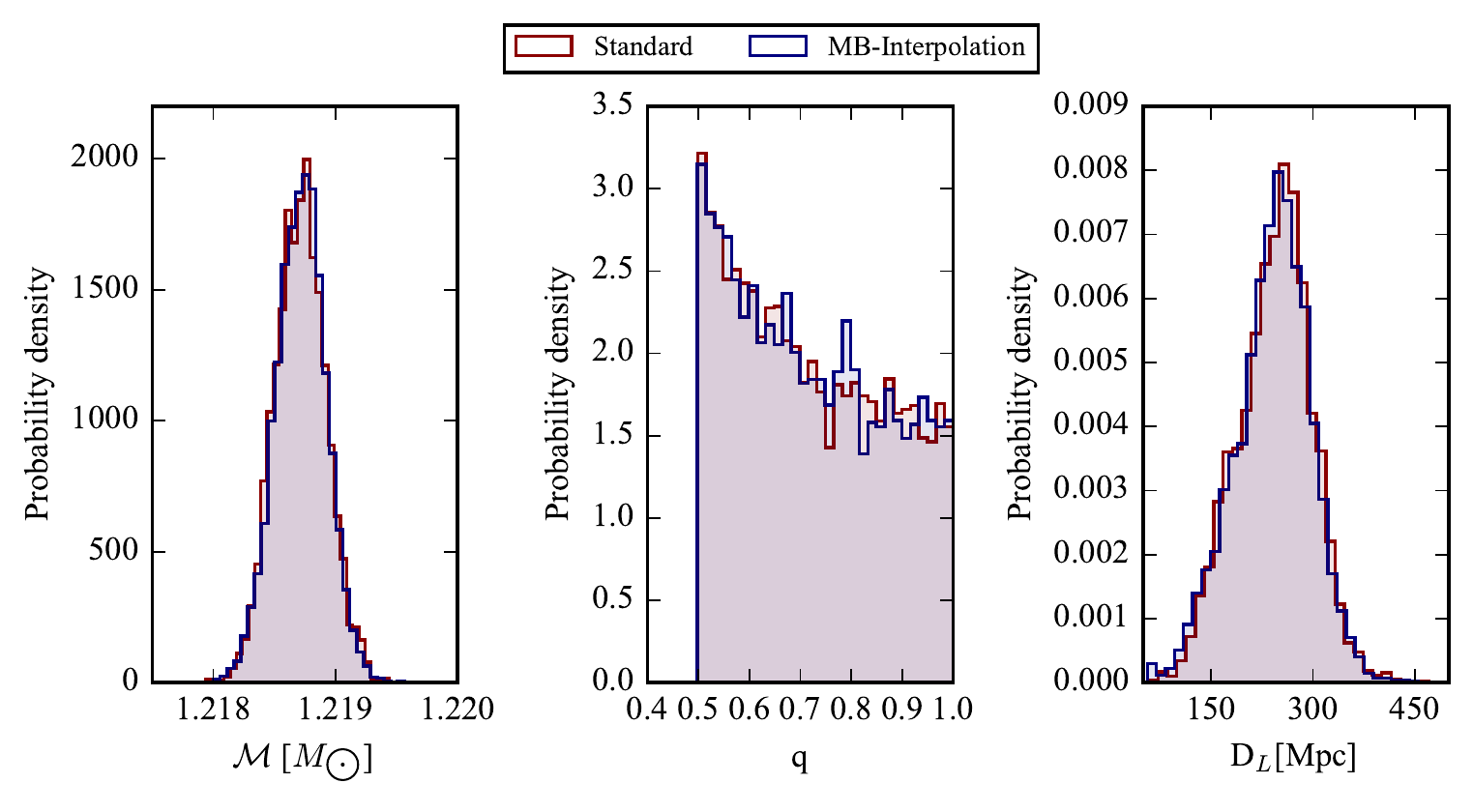}
 \caption{Posterior distributions of chirp mass $\mathcal{M}$, mass ratio $q$ and luminosity distance $D_L$ with the {\it TaylorF2} waveform model.  \label{Distrib}}
\end{figure}

The analysis of the mock data with MB-Interpolation templates 
produced posterior distributions statistically identical to the ones obtained with a  
standard analysis.
As a representative case, in figure \ref{Distrib} we show the marginalized posterior probability density functions for chirp mass, mass ratio and luminosity distance,
the quantities most sensitive to phase and amplitude errors.  We confirmed the visual agreement 
between the marginal probability distributions obtained with the standard and the MB-Interpolation algorithms by performing a Kolmogorov-Smirnov test, which showed that the two sets of samples are consistent with random draws from the same distribution.

\subsubsection{PE speedup}
\begin{figure}
\begingroup
    \centering
    \includegraphics[]{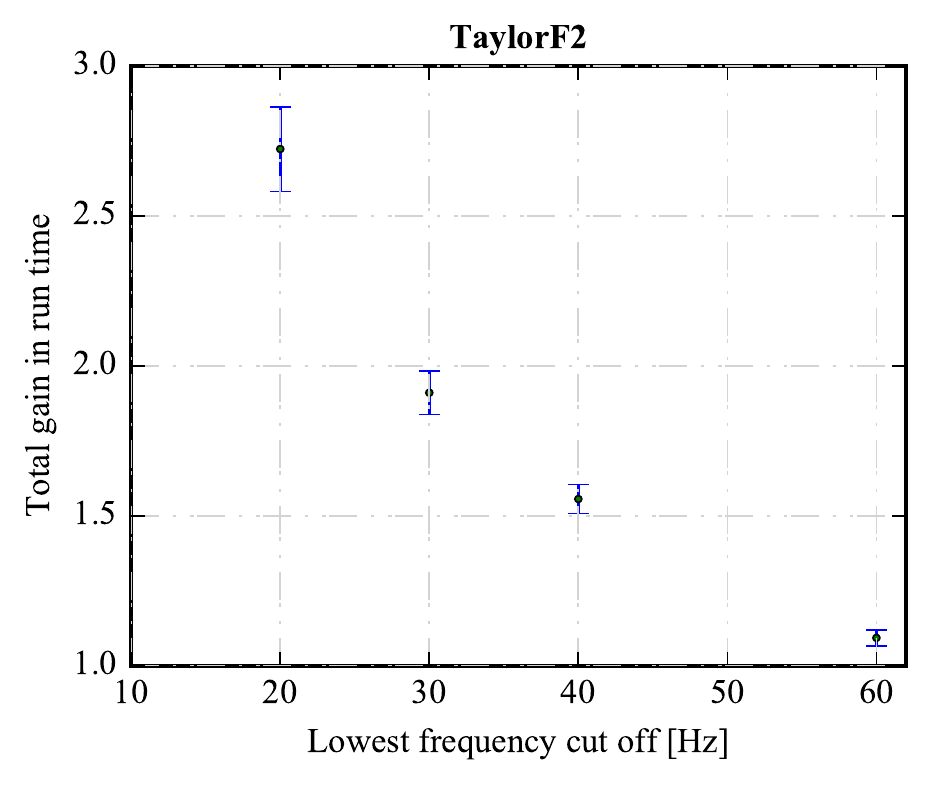}
    \caption{Observed gains in the end-to-end PE run-time, ${G_\mathrm{PE}^\mathrm{TF2}}$, as a function of the starting frequency for analyses carried out with {\it TaylorF2} waveforms. \label{fig_6} 
    }
\endgroup
\end{figure}

To test the effect of the speedup in the waveform generation on the overall PE analysis, we measured the computational time required to perform end-to-end PE runs. 
We performed PE analyses from different values of the starting frequency $\fmin$, consequently changing the lengths of the data segments.  The results are reported in table \ref{tabGain} and
figure \ref{fig_6}. 

For each starting frequency and for both standard and MB-Interpolation algorithms, the times of 4 runs have been averaged.
The ratio between the average time required to complete a PE analysis adopting the standard and the MB-Interpolation algorithms has been used to define the overall speedup gain $G_\mathrm{PE}$.
Each group of 4 identical analyses has been run at the same time on a Dual-Core AMD Opteron 2218 Processor with a clock speed of 2.6GHz.

Table \ref{tabGain} reports the measured speed gain for the whole PE analysis with the {\it TaylorF2} (abbreviated as TF2) waveform model in the third column. 
The fourth column contains the speedup in the template calculation (cf.~Fig.~\ref{fig2_2}). In the last two columns we also report the ratio between the number of frequencies at which the waveform is evaluated in the standard and the MB-Interpolation algorithms $\Nfix/N_\mathrm{MB}$, as well as the idealized improvement in the limit of continuously adapted sampling rates $\Nfix/N_\mathrm{min}$ (equation \ref{eq:idealG}).

As can be seen from table \ref{tabGain}, we do not obtain the full idealized gain that may be expected from multi-banding for three reasons.  Firstly, the actual number of frequencies at which the waveform is computed in our multi-banding algorithm is larger than the theoretical limit, so $\Nfix/N_\mathrm{MB} < \Nfix/N_\mathrm{min}$.  Secondly, the template computation speedup is less than the reduction in the set of frequencies due to multi-banding, ${G_\mathrm{template}} < \Nfix/N_\mathrm{MB}$, because of the additional cost of interpolation.  Thirdly, the PE speedup is smaller than the speedup in template generation, ${G_\mathrm{PE}}<{G_\mathrm{template}}$, because template generation is only one component of the PE algorithm.
Although the waveform computation is the dominant computational cost for computationally expensive templates, the cost of evaluating the likelihood still grows with the number of frequency bins even when using MB-Interpolation, and along with interpolation this can become the most expensive step when using MB-Interpolation with very long waveforms.

We did not repeat the end-to-end parameter estimation calculations across the full range of starting frequencies with {\it IMRPhenomPv2} waveforms because the computational cost was unacceptably high when using the standard procedure.  Nonetheless, it is possible to estimate the computational cost gain one would achieve with {\it IMRPhenomPv2} waveforms when starting with low values of $\fmin$.  For computationally expensive waveforms, the parameter estimation cost is dominated by the waveform computation cost; this is a factor of $\sim$ 3 higher for {\it IMRPhenomPv2} waveforms than for {\it TaylorF2} waveforms.  (This factor is independent of the waveform duration or starting frequency, and reflects the difference in the cost of computing the two waveforms at a given frequency point.)  Therefore, we expect that the total PE computational cost with {\it IMRPhenomPv2} waveforms to be about the same factor of 3 larger than for {\it TaylorF2} waveforms when starting at low frequencies and using the standard procedure.
Meanwhile, for sufficiently low starting frequencies, the MB-Interpolation waveform computation cost is dominated by interpolation, so that the template computation cost with {\it IMRPhenomPv2} and {\it TaylorF2} waveforms when using MB-Interpolation asymptotes to the same value -- and so does the full PE computational cost.  Thus, we expect that end-to-end PE gains from using MB-Interpolation with {\it IMRPhenomPv2} will be a factor of $\sim 3$ greater than with {\it TaylorF2} waveforms.

MB-Interpolation is most effective for smaller values of the starting frequency. Fig.~\ref{fig2_2} shows that with $\fmin=8$\,Hz the template speed-up factor is $\sim 16$ for {\it TaylorF2} waveforms and $\sim 50$ for {\it IMRPhenomPv2} waveforms, {where we conservatively assume that there are no further significant gains in waveform computation at lower starting frequencies because of fixed interpolation costs.  As discussed above, the total PE speed-up will not be as large as the speed gain in template generation because of other fixed costs.  Nevertheless, as next generation interferometers (such as the Einstein Telescope \cite{ETDesign2011}, KAGRA \cite{aso2013interferometer} and the Cosmic Explorer \cite{LIGOinstrument2015}) take advantage of low-frequency data, MB-Interpolation should improve parameter estimation costs by factors of tens, or more for more expensive waveform models.}  

\begin{table}[hbt!]
\centering
\begin{tabular}{|c|c|c|c|c|c|}
\hline
$\mathbf{\fmin} [Hz]$ & $\boldsymbol{\delta}\mathbf{f_\mathrm{0}}[Hz]$ &  $\mathbf{G_\mathrm{PE}^\mathrm{TF2}}$& $\mathbf{G_\mathrm{template}^\mathrm{TF2}}$ & $\mathbf{\Nfix/N_\mathrm{MB}}$ & $\mathbf{\Nfix/N_\mathrm{min}}$\\
\hline
 $60$ & $1/16 $ & $1.09 \pm 0.03$ & $ 1.31\pm 0.01$ & $3.76$ & $55.4$ \\
\hline
 $40$ & $1/64$ & $1.56 \pm 0.05$ & $3.8\pm0.1$ & $12.82$ &  $83.8$ \\
\hline
 $30$ & $1/128$ &  $1.91 \pm 0.07$ & $5.5\pm 0.1$ & $23.40$& $112.2$ \\
\hline
 $20$ & $1/300$ &  $2.72 \pm 0.14$ & $8.8 \pm 0.2$ & $61.01$ & $169.1$ \\
\hline
\end{tabular}
\caption{The table reports the results obtained for different values of starting frequency $\fmin$ (first column) and corresponding sampling steps $\delta f_\mathrm{0}$ for standard template generation (second column). The values $G^\mathrm{TF2}_\mathrm{PE}$  are the actually measured speed gains in the complete PE analyses with {\it TaylorF2} due to using MB-Interpolation.  $G^\mathrm{TF2}_\mathrm{template}$
is the gain in the waveform generation speed
($T_{\mathrm{Standard}}/T_{\mathrm{MB-Int.}}$). 
$\Nfix/N_\mathrm{MB}$ is the ratio between the number of frequencies at which the standard and multibanded waveforms are evaluated, while $\Nfix/N_\mathrm{min}$ is the limiting case for the reduction in waveform evaluations when continuously adapting frequency steps. 
\label{tabGain}}
\end{table}

\section{Conclusion}\label{s:conclusion}
Parameter estimation has played an important role in the opening of the field of gravitational wave astronomy, as demonstrated in the analysis of Advanced LIGO's first observations~\cite{GW150914-PARAMESTIM,CBC-O1-BBH}.
The stochastic sampling algorithms used in these analyses require the generation of millions of template waveforms which are compared to the data, a computational task that becomes more expensive as the in-band signal duration increases: for signals from lower mass binaries and for detectors with improved sensitivities at lower frequencies. The generation of computationally expensive template waveforms is the bottleneck in the PE analysis, limiting our ability to obtain results quickly. In this paper we proposed an alternative approach to reduce this cost and consequently the overall time required to produce a result.
The procedure is inspired by the same {\it multi-banding} approach already adopted for low-latency algorithms dedicated to gravitational wave searches~\cite{cannon2012interpolating,adams2015low}.
It consists in reducing the set of frequencies at which to evaluate waveforms by dividing the spectral range into different bands and optimising the sampling procedure. However, the greater accuracy required in the context of PE demanded an additional up-sampling of the waveform when computing the likelihood function, which led us to apply a linear interpolation in phase and amplitude.

We have demonstrated the effectiveness of the method by implementing it in the \texttt{LALInference} PE code and comparing the results to inference with the full waveform.  We found negligible differences between the results at a greatly reduced computational cost.
We showed that the MB-Interpolation algorithm reduces the number of frequencies 
at which the waveform is evaluated by more than an order of magnitude for $\fmin< 40$\,Hz. This leads to an acceleration of the waveform generation and, consequently, the whole analysis. For a fixed chirp mass of the binary, the speed-up factor depends on the complexity of the model and on the starting frequency \fmin. We studied the most challenging case of binary neutron stars, adopting the {\it TaylorF2} and {\it IMRPhenomPv2} waveform families. In section \ref{s:results} we reported speedup factors in the template generation which reached {$\sim 50$ for the most sophisticated waveform model ({\it IMRPhenomPv2}) at $\fmin \sim 10$\,Hz.  Although the overall decrease in the computational cost of end-to-end PE is more modest than the improvement in template generation because of fixed costs, we expect factors of tens in speed gain when using {\it IMRPhenomPv2} templates with starting frequencies of a few Hz.
The considerable speedup gains reached by the implementation of the MB-Interpolation method demonstrates the effectiveness of the approach.

Our method is related to the reduced order quadrature models of gravitational waveforms introduced for the simple TaylorF2 model in \cite{Field:2011} and later created for more sophisticated SEOBNR and IMRPhenomP models~\cite{Smith:2016qas}.
These methods also result in a large acceleration of PE, by factors of $70$ for a TaylorF2 waveform ~\cite{Canizares:2014} or $300$ for IMRPhenomPv2~\cite{Smith:2016qas} from 20\,Hz.
The two methods are conceptually similar in that the number of points in frequency at which the waveform is evaluated is reduced.  However, for ROQ the interpolation makes use of a different set of bases, which means that interpolation must be performed across the parameter space of the signals in addition to interpolation in frequency. Unlike our MB-Interpolation method, this requires significant setup costs to create the parameter space interpolants, which are difficult to produce for the higher dimensional precessing spin parameter space. In this regard our method is more flexible, since it can be used for any signal without pre-computing a reduced order model. This is advantageous in the case of models which include additional physical parameters, such as neutron star tidal deformability, which further increase the dimensionality of the parameter space.
The MB-Interpolation method can also be used in combination with ROQ, where the MB-Interpolation is used to accelerate the initial creation of the ROQ model by reducing the number of calculations and also the memory overhead. Indeed, the reduced frequency basis idea was used in \cite{Smith:2016qas}, but without the interpolation up-sampling step.

Accelerated waveform generation techniques such as the one we have developed here are likely to be essential in future, as the detectors evolve toward greater sensitivity at low frequencies. Third-generation detectors such as the Einstein Telescope~\cite{ETDesign2011} or LIGO Voyager~\cite{3GLIGO} will be sensitive down to a few Hz, meaning signals may be in band for hours or longer.
The same MB-Interpolation procedure can also be  applied to GW studies in the context of space missions \cite{porter2014new}, and in particular for phase-coherent modeling of the signal in both space-based and ground-based detectors as would be useful for joint science exploitation~\cite{PhysRevLett.116.231102,PhysRevLett.117.051102}.

{In principle, a similar multi-banding approach could also be applied to time-domain
waveforms, which could be sampled at a lower rate earlier in the waveform. However, the time domain waveforms of greatest interest use numerical integration of the waveform with an adaptive step size in time. This prevents a great speedup from being obtained as one cannot reduce the step size arbitrarily in the simple way we could for the frequency domain waveforms. This factor, in addition to the technical difficulty of efficiently reconstructing the FFT of a non-uniformly sampled time series prevented us from exploring this option in the current work.}

Finally we note that despite the specialisation to gravitational wave analysis, the same technique of adapting the sampling interval could be applied to any area where the signal frequency changes monotonically with time.
 
\section*{Acknowledgements}
We thank Rory Smith for useful comments on an earlier version of this manuscript. The research leading to these results has received funding from the 
People Programme (Marie Curie Actions) of the European Union's Seventh 
Framework Programme FP7/2007-2013/ (PEOPLE-2013-ITN) under REA grant 
agreement no.~[606176]. This paper reflects only the authors' view and the European Union is not liable for any use that may be made of the information contained therein. JV was supported by UK Science and Technology Facilities Council (STFC) grant ST/K005014/1. IM was partially supported by the STFC.

\section*{Bibliography}
\bibliographystyle{unsrt}
\bibliography{bibliography.bib}
\end{document}